\documentstyle[11pt,aaspp4, flushrt]{article}  

\begin{document}
 
\title{The Composite Broad Band Spectrum of Cir X--1 at the Periastron: 
a Peculiar Z--source}

\author{R. Iaria\altaffilmark{1},
L. Burderi\altaffilmark{2},
T. Di Salvo\altaffilmark{1},
A. La Barbera\altaffilmark{1},
N. R. Robba\altaffilmark{1}}
\altaffiltext{1}{Dipartimento di Scienze Fisiche ed Astronomiche, 
Universit\`a di Palermo, via Archirafi n.36, 90123 Palermo, Italy}
\altaffiltext{2}{Osservatorio Astronomico di Roma, Via Frascati 33, 
00040 Monteporzio Catone (Roma), Italy}
  
\authoremail{iaria@gifco.fisica.unipa.it}

\begin{abstract}

We report on the  spectral analysis
of the peculiar source Cir X--1 observed by the BeppoSAX satellite when the
X--ray source was near the periastron.
A flare lasting $\sim 6 \times 10^3$ s is present at the beginning of the 
observation. 
The luminosity during the persistent emission  is $1 \times 10^{38}$ erg 
s$^{-1}$, while during the flare is $2 \times 10^{38}$ erg s$^{-1}$.
We produced  broad band (0.1--100 keV) energy spectra during the flare and
the persistent emission. At low energies the continuum is well fitted  by a 
model consisting of Comptonization of soft photons, with a temperature
of $\sim 0.4$ keV, by electrons at a 
temperature of $\sim 1$ keV.  Out of the flare  a power-law component, 
with photon index $\sim 3$, is dominant at energies higher than 10 keV. 
This component  contributes to 
$\sim 4\%$ of the total luminosity. During the flare its addition 
is not statistically significant.  
An absorption edge at $\sim 8.4$ keV, with optical depth  $\sim 1$, 
corresponding to the K-edge of Fe XXIII--XXV,
 and an iron emission line  at 6.7 keV are also present. 
The iron line energy is in agreement with the ionization level 
of the absorption edge. The  hydrogen column deduced from  the absorption 
edge is $\sim 10^{24}$ cm$^{-2}$,
two order of magnitude larger than the  absorption measured in this 
source.  
We calculated  the radius of the region originating the comptonized 
seed photons, $R_W \sim 150$ km.
 We propose  a scenario where $R_W$ is the inner disk radius,  a highly 
ionized torus  surrounds
the accretion disk and a magnetosphere is large up to  $R_W$.   
The absorption edge and the emission line  could 
originate in the photoionized  torus, while the comptonized component  
originates in an inner region of the disk. 
\end{abstract}


\section{Introduction}

Circinus X--1 (Cir X--1)  is one of the most enigmatic X--ray binary sources 
and shows a complex temporal and spectral behaviour. The earliest 
unambiguous X--ray detection was made by Margon et {\it al.} (1971). 
The position of the  probable radio 
counterpart is located near the supernova remnant G321.9--0.3 (Clark,
Parkinson \& Caswell, 1975), 
suggesting that Cir X--1 could be  a runaway binary.
The source displays a periodicity at radio (Whelan et {\it al.}, 1977), IR 
(Glass, 1978), optical (Moneti, 1992) and X--ray wavelengths (Kaluzienski et 
{\it al.} 1976) with a period of $\sim 16.6$ days.
The high degree of stability of this period  
suggests that it is the orbital period. Periodic radio flares near the
zero phase (for recent ephemeris see Stewart et  {\it al.}, 1991) of the
orbital period are also
accompanied by drastic changes in the X--ray light curve (X--ray flares
and dips).
These variations  can be explained in terms of a 
highly eccentric binary orbit, where the periastron, and consequently the 
increased mass transfer, occur near zero phase (Murdin et {\it al.} 1980).
Shirey (1998) observed  that the
dipping activity of Cir X--1 occurs within 0.03 and 0.09 in the phase interval
when the source enters its flaring
state,  after  an increase of intensity in the X--ray light curve (Brandt et
{\it al.}, 1996;   Shirey et {\it al.}, 1996).
These dips  might be explained by obscuration due to 
inhomogeneities in the mass transfer stream or intermittent bumps
on the outer edge of the accretion disk (Shirey, 1998).  
There is evidence  of presence of radio jets from the
source (Stewart et  {\it al.}, 1993).

Cir X--1 shows a long-term variability  in its luminosity. 
EXOSAT observations show  a gradual increase in the
brightness of the source near zero phase from 1984 to 1986, 
models involving the precession of the accretion disk, the thickening of
the outer disk, or an apsidal motion have been proposed (Oosterbroek et 
{\it al.}, 1995). 

The optical counterpart of Cir X--1 is a faint red star (Moneti, 1992). 
The large ratio $L_x/L_{opt}$ suggests that   Cir X--1 is a
low--mass X--ray binary
(LMXB). For a long time it was suspected that the compact object in Cir X--1
 was a black hole, for the rapid variability similar to that of Cyg X--1 
(Toor, 1977). However the presence of a neutron star was demonstrated by the 
detection of type I X--ray burst (Tennant et {\it al.}, 1986 a,b). 
The spectral and fast timing properties of accreting low magnetic field
neutron stars permit to distinguish them into two class,  
Z and Atoll sources, from the pattern each source describes 
 in a X--ray color-color diagram (Hasinger \& van der Klis, 1989). 
The Z--sources are very bright and could have a  moderate  magnetic field,
 while 
the Atoll--sources have a lower luminosity and probably a weaker magnetic 
field.
It was not clear 
what kind of LMXB  Cir X--1 is. Oosterbroek et {\it al.}
 (1995) proposed that Cir X--1 may be an 
 Atoll--source with a magnetic field strength of less than $10^9$ Gauss.      
Using RXTE data  Shirey (1998) identified much of the behaviour of Cir X--1 
as that of a 
Z--source: hardness-intensity diagrams showed that the track differs somewhat
 from the canonical Z track, but the temporal behaviour of the source 
along the  branches strongly suggests that Cir X--1 is a Z--sources.
The presence of horizontal--branch oscillations, in the range 1--20 Hz, 
 in the framework of the beat-frequency 
model (Alpar \& Shaman, 1985; Lamb et {\it al.}, 1985), could indicate a 
significant magnetic field.
Quasi periodic oscillations (QPO) at 100--200 Hz have
been reported (Tennant, 1987; Shirey et {\it al.}, 1996; Shirey, 1998),
but no kHz QPOs were detected.

The distance to Cir X--1 was estimated to be between 8--10 kpc (Goss \&
Mebold, 1977). Recently, Case \& Bhattacharya (1998) have revised the 
 distance to the supernova remnant G321.9-0.3, and hence  to the associated 
Cir X--1, to 5.5 kpc. In the following we adopt 5.5 kpc as the distance to 
the source, unless otherwise stated.

The spectrum of Cir X--1 has been poorly understood up to now. ASCA data 
taken near the zero phase, at a low count rate, have been fitted with an
absorbed two--blackbody model (Brandt et {\it al.}, 1996). An iron K edge at 
7 keV was present,  
that corresponds  to a large absorption column of $\sim 10^{24}$ cm$^{-2}$
 and can be explained by 
 a partial covering model. A weak emission line at 6.4 keV  has also
been  detected. The edge and the gaussian emission line were no
longer detected at higher  count rate, above 300 c s$^{-1}$ 
(Brandt et {\it al.}, 1996).
In a recent RXTE observation (Shirey et  {\it al.}, 1999a) the spectrum of  
Cir X--1 was fitted using the Eastern Model (Mitsuda et  {\it al.}, 1984), 
i.e. a  multicolor disk blackbody,
with a temperature between 1.2--1.4 keV, 
and a  blackbody  with a temperature in the interval 1.9--2.4 keV. 
They found  a peculiar feature, very prominent in the flaring branch,
around 10 keV not easily
understandable. This feature could be fitted with an absorption edge, but even 
a narrow line improved the fit.

Here we report  the results of a broad band (0.1--100 keV) spectral analysis
performed on Cir X--1 data from the BeppoSAX satellite.  The broad 
energy coverage of BeppoSAX narrow field instruments allows us to well
 constrain the spectrum of this source, and to detect the presence of a 
high energy component.

\section{Observations and Light Curve}

The Narrow Field Instruments (NFI) on board BeppoSAX satellite
(Boella et al. 1997) observed Cir X--1 from 1998 August 1 15:59:54.5 UT, 
for a total time of 90 ks.  
The NFIs are four co-aligned instruments which cover more
than three decades of energy, from 0.1 keV up to 200 keV, with good
spectral resolution in the whole range. LECS (operating in the range 
0.1--10
keV) and  MECS (1--11 keV) have
imaging capabilities with a field of view of $20'$ and $30'$ radius, 
respectively. We selected the data for scientific analysis 
in circular regions, centered on the source, of $8'$ and $4'$ radii for  
LECS and MECS respectively. The background subtraction
was obtained using blank sky observations in which we extracted the 
background spectra in regions of the field of view similar to those used for 
the source. HPGSPC (7--60 keV) and PDS (13--200 keV) have not imaging 
capabilities, because their field of views, of $\sim 1^\circ$ FWHM, are 
delimited by collimators. The background subtraction for these instruments is 
 obtained using the off-source data accumulated during the rocking 
of the collimators.    
The energy ranges used in the spectral analysis are: 
0.12--4 keV for LECS, 1.8--10.5 keV
for  MECS, 7--25 keV for  HPGSPC  and 15--100 keV for  PDS.
LECS was only active  in the time interval between
$3.4 \times 10^4$  s and $\sim 5.9 \times 10^4$ s from the beginning of the 
 observation. 
Different normalizations of the NFIs are considered 
by including in the model normalizing factors, fixed to 1 for  MECS, and 
kept free for the other instruments. 
A systematic error of $2\%$ was added to the spectra.
Assuming the orbital ephemeris as reported by Stewart et {\it al.} (1991), 
we find that our observation covers  the phase interval from 0.11 to 0.16. 

In Figure 1 we plotted the  Cir X--1 light curve in the energy band
1.8--10 keV (MECS data, upper panel), and the  ratio of the 
count rate in the energy band 3--7 keV  to that in the  
band 1--3  keV (lower panel). No dips are present in the light curve.
A flare that lasts $\sim 6 \times 10^3$ s is present.
 The average count rate, out of the flare, is 
$\sim 300$ counts/s in the MECS, corresponding to an average hardness ratio of 
$\sim 0.6$. 
The count rate increases during the flare, reaching  $\sim 700$ counts/s
at the peak, where the hardness ratio is  $\sim 0.9$. We also 
observe that the hardness ratio  increases 
from the end of the flare to the end of the observation (see Fig. 1).
We plotted in Figure 2 the color-color diagram (hereafter CCD) 
of the source  using
a bin time of 137 s. The variations  are mainly due 
to the presence of the flare.
The shape of this CCD is reminiscent of the flaring branch observed in
the Z--Sources.
We divided our observation
in 8 intervals as reported in Table 1 and in Fig. 1. The first five 
intervals are
taken during the flare in order to
study the rapid spectral variation. The last  three  intervals
are taken out of the flare and correspond to different 
hard colors and  positions in the CCD.

In Table 1 we show the flux of the source in the 2.5--25 keV energy range
for all the intervals in order to compare these  values with previous
observations.
In the same table we report the average observed flux of the source
in the 0.1--100 keV energy range,
and the corresponding luminosity of the source.
The LECS data are not always available; in Table 1 we report for which 
intervals  LECS was active.

\section{Spectral Analysis}
  
As described above we extracted five spectra during the flare and three 
spectra during the persistent emission corresponding to different hard colors.
The LECS was active only in two of the eight intervals (see Tab. 1). 
The average photoelectric absorption by cold matter,  
obtained fitting the spectra containing the low energy data, is
$\sim 1.74 \times 10^{22}$ cm$^{-2}$.
This parameter was fixed in the other intervals. A simple model,  
such as  a blackbody or a disk multicolor  blackbody, plus  
photoelectric absorption by cold matter, is not sufficient to fit the
data. 
The spectra during the flare at higher count rate (intervals from 2 to 4) 
and the soft part (up to 10 keV)
of all the other spectra are well fitted by a comptonized model 
({\it Comptt}, Titarchuk 1994), the Comptonization of
soft seed photons in a Wien distribution by hot electrons.
We obtained  seed-photon and electronic 
temperatures of $\sim 0.4$ keV  and  
$\sim 0.9$ keV, respectively, and both seem to increase during the flare 
(see Fig. 4). 
The Compton optical depth is around 17 for all the spectra. 
A blackbody  plus a disk multicolor  
blackbody give  equivalent fits. In this case  we obtain a  disk
temperature of $\sim 0.4$ keV, similar to the seed photon temperature 
and a blackbody temperature of $\sim 0.9$ keV, corresponding to the electronic
 temperature of the Comptonization model.

The spectra at lower count rate 
(intervals 1, 5, 6, 7, 8) show a marked hard excess above 
10 keV with respect to the previous models. The addition 
of a  power law to the model improves 
the fit with very high statistical significance, although  
 a thermal-bremsstrahlung component instead of the power law  can not be 
excluded. The power-law 
component contributes to $\sim 4\%$ of the total luminosity and its    
photon index is between 3.17 and 3.37.    
We also added the
hard power-law component  to the other spectra at high count rate
 (intervals 2, 3, 4), fixing 
the photon index to the average value 3.3. In these cases, the power-law
component does not improve the fit, and we  obtain an upper limit 
on the normalization.
In Table 2 and 3 
we report the result of the fits for all the intervals and the
probability of chance improvement for the addition of a power law component. 

A feature between 8 and 10  keV is clearly visible in all the spectra 
and was also present in previous observations (e.g. Shirey et {\it al.}, 
1999a). 
We add an edge at $\sim 8.4$ keV to fit this feature improving the fits 
with high statistical significance. 
In spectra 6, 7, 8 the addition of an iron line improves furtherly the 
fit (the addition is significant  at a confidence level 
between  99.8\%  and 99.996\%). 
The iron line energy is $\sim 6.7$ keV, with  an upper limit for sigma of 
0.24 and 0.47 keV  (spectra 6 and 7 respectively), and a sigma of 0.5 keV for 
spectrum 8. We find that the equivalent widths are between 46 and 100 eV
(see Table 3).
In Figure 3 we report the spectra and the residuals in units of $\sigma$ with
respect to the best fit model for  intervals 4 (at the
peak of the flare) and 6 (out of the flare).
In Figure 4 we report the unfolded spectrum for the same intervals 
shown in Figure 3.

\section{Discussion}

We analyzed data of Cir X--1 from a BeppoSAX 90 ks
observation in the energy range 0.1--100 keV.  The light curve  
shows a flare lasting $\sim 6 \times 10^{3}$ s at the beginning  of the 
observation and does not show dipping activity.  
The CCD of the source that we obtained is similar to the FB of a typical
Z--source, in  line with RXTE results (Shirey et  {\it al.}, 1998; 1999a).

The equivalent absorption column $N_H$ is $ \sim 1.74 \times 
10^{22}$ cm$^{-2}$, and is in agreement with previous observations (Brandt
 et {\it al.}, 1996). 
For a distance to the source of 5.5 kpc  (Case \& Bhattacharya, 1998) 
the visual extinction in the 
direction of Cir X--1 is $A_v=5.8 \pm 2.0$ mag (Hakkila et al. 1997). 
Using the observed correlation between visual extinction and 
absorption column (Predehl \& Schmitt 1995) we find 
$N_H=\left(1.38 \pm 0.02 \right) \times 10^{22}$ cm$^{-2}$. 
It is  slightly smaller than the value obtained from the fit,
probably because of  the presence
of obscuring matter close to the X--ray source.

A comptonized spectrum  with  an iron absorption edge are needed to fit 
the broad band energy spectra.
For the spectra with lower count rate (intervals 1, 5, 6, 7, 8)  
the addition of an
extra hard component is needed to fit the high energy range.
The temperatures of the seed photons
and the hot electrons of the comptonized component increase during the flare. 
In Figure 5 we plotted the two temperature versus the  time
 (centers of the corresponding temporal intervals).
The optical depth of the comptonizing region is $\sim 17$ and 
does not seem to vary. 
The unabsorbed luminosity of comptonized  component is  $\sim 2 \times 10^{38}$
erg s$^{-1}$ during the persistent emission, and  reaches
 $\sim 3.5 \times 10^{38}$ erg s$^{-1}$  at the peak of the flare.
In Figure 6 we report the electronic temperature  versus the seed-photon 
temperature.  
A  linear correlation between the two temperatures is
evident, with a slope of $3.2 \pm 1.2$. 
This correlation could indicate a sort of equilibrium 
between the two temperatures, because an increase of the seed-photon 
temperature corresponds to an increase of the electron temperature.    
The comptonizing region  is probably located in the inner part of the system,
covering  part of the accretion disk, where the seed photons could originates. 

An Eastern model (blackbody plus multicolor disk blackbody) instead of a
comptonized model also fits  well the data. The 
inner radius of the disk that we obtain in this case is $\sim 160$ km, assuming
an inclination angle of $60^{\circ}$, similar to the radius of the seed 
photons obtained from the previous model (see below).  
On the other hand the radius of  $\sim 35$ km obtained from the blackbody
is  quite large to be identified with the  neutron star radius.  
 It is possible that the blackbody mimics the Wien law of the Comptt 
model. Because  the Comptt model seems more reasonable we adopt this 
interpretation.

\subsection{Possible presence of a magnetic field of $\sim 4 \times 10^{10}$
 Gauss} 

For each interval we calculated the radius of the seed-photon Wien spectrum, 
 $R_W$, using the parameters reported 
in Tables 2 and 3. Following in 't Zand et {\it al.} (1999) 
 this radius can be expressed as 
$R_W=3 \times 10^4 D \sqrt \frac{f_{bol}}{1+y}/ \left(kT_0 \right)^2$ km,
where $D$ is the distance to the source in kpc, 
$f_{bol}$ is the unabsorbed flux of the Comptt
model in erg cm$^{-2}$ s$^{-1}$, $kT_0$ is the seed photon 
temperature in keV, and  $y=4kT_e\tau^2/m_ec^2$ is the relative energy gain 
due to the Comptonization.  
  We 
obtain that  the  seed-photon radius is $\sim 150$ km.   
An hypothesis is that this is the inner  radius of the accretion disk.
It is quite large but we observe that a large inner disk radius is in 
agreement with the absence of kHz QPOs in the 
power spectrum density  of Cir X--1 (Shirey et {\it al.}, 1996).
In fact kHz QPOs are usually explained as the keplerian frequency at the inner
edge of the disk,
near the neutron star surface (Strohmayer et al.  1996; Miller, Lamb, \& 
Psaltis 1998; Stella \& Vietri 1999; Titarchuk, Lapidus, \& Muslimov 1998). 
The lack of kHz frequencies in the power spectra of Cir X--1
could mean that the inner region of the disk is not present.     

A  possible explanation  of the large value of the inner radius
is that the compact object presents
a magnetic field sufficiently strong to obtain a magnetosphere around
the neutron star, although in this case we need to suppose a particular 
geometrical configuration in order to avoid the unobserved coherent 
pulsations. 
The relation between the magnetospheric radius $R_m$ and the 
magnetic field strength is 
$R_m=4.3 \times 10^3 \phi \mu_{30}^{4/7}R_6^{-2/7}L_{37}^{-2/7}\epsilon^{2/7} 
m^{1/7}$ 
km (see e.g. Burderi et {\it al.}, 1998), where $\phi \sim 0.5$ 
is a correction factor for the Alfven radius when an accretion disk
is present (Ghosh \& Lamb, 1991), 
$\mu_{30}$ is the magnetic moment in 
units of $10^{30}$ G cm$^3$, $R_6$ is the neutron star radius, $R_{NS}$,
in units of 
$10^6$ cm, $ \epsilon=GM \dot{M}/R_{NS}$ is the ratio between the observed 
luminosity and the 
total gravitational potential energy released per second by the accreting
matter, $m$ is the neutron star mass in unit of 
solar mass. Adopting $R_6=1$ and $m=1.4$ (typical for a neutron star),  
$ \epsilon \sim 1$, $L_{37}$ is the intrinsic luminosity of the 
comptonized component in units of  $10^{37}$ erg s$^{-1}$,  and 
$R_m=R_W$ we can find  a magnetic field strength for each interval. 
The obtained values are consistent with an average magnetic field of  
 $\sim 4 \times 10^{10}$ Gauss.
This value is slightly higher than the typical magnetic field strength 
inferred for the Z--sources.

Another possibility to explain the large inner radius of the accretion disk
is that the inner region of the disk is hidden for the 
presence of optically thick material (e.g. the comptonized region with 
Compton optical depth $\sim 17$) in the central part of the system.

\subsection{Detection of a hard tail in the energy spectra}

A hard component is required to fit the spectra at low luminosity 
 (1, 5, 6, 7, 8). During the flare 
 the hard component is not detected with statistical significance 
(see Tab. 3, where the upper limit for the normalization is reported).
A similar hard component has also been observed in GINGA data of 
GX 5--1 (Asai et {\it al.}, 1994), in BeppoSAX data of GX 17+2 
(Di Salvo et {\it al.}, 2000) and in BeppoSAX data of GX 349+2 (Di Salvo 
{\it al.}, 2000b).
Asai et {\it al.} (1994) observed  GX 5--1 in the  normal branch (NB) and in 
the flaring branch (FB) of the relative Z-track; the intensity of the power-law
component decreases from the NB to the FB.
Di Salvo et {\it al.} (2000)  observed  GX 17+2 in the horizontal branch (HB)
and in the NB, and 
the intensity of the power-law component was found to decrease toward the 
NB/FB vertex.
 In Cir X--1 the intensity of the power-law component decreases from the
NB/FB vertex  toward the FB.

The hard excess could have a non thermal origin, due to scattering of
soft photons by 
electrons with a non thermal distribution of velocities. These electrons 
could be part of a jet. In fact Stewart et {\it al.} (1993) detected 
the presence
of a jet with a maximum at the phase 0.17 where the flux 
 increased up to 12 mJy and  
an upper limit on the 12-h integrated flux density of 
$\sim 3 \pm 1 $ mJy.  
The angle between the  jet and the line of sight is $> 70^{\circ}$ 
(Stewart et {\it al.}, 1993; Fender  et {\it al.}, 1998).  
Our observation is in the phase interval   0.11--0.16, near the observed 
maximum of the radio flux. 
So, a jet could  be the origin of the power law observed in Cir X--1,
and similarly in GX 5--1 and GX 17+2. These  sources, as well as all the
Z--sources,  have been 
associated with variable radio sources (Fender \& Hendry 2000, 
and reference therein).  Note that these hard 
component might be a general feature of the Z-sources and this is another 
feature that Cir X--1 shares with the Z--source family.         
     
\subsection{Could an ionized torus be present around the disk?}
An absorption edge at $\sim 8.4$ keV is present in all the spectra, 
and a gaussian emission line at $\sim 6.7$ keV is
detected in the spectra 6, 7 and 8 (see Tab. 2 and 3).
The energy of the emission line is
compatible with the iron ionization levels Fe XXIII--XXV indicated by 
the absorption edge energy (see Turner et {\it al.}, 1992 for a correspondence
 between  iron edge energy and ionization level).
The best fit value for the optical depth $\tau_{edge}$, considering the 
photoionization cross section for the K--shell Fe XXIII 
(Krolik \& Kallman, 1987), corresponds to a 
hydrogen column density of $\sim 1.5 \times 10^{24}$ cm$^{-2}$, assuming a 
cosmic abundance of iron. This is much higher than the measured photoelectric
absorption (see Tab. 2 and 3).  This could 
imply the existence of a highly ionized region  around the compact
object, producing the absorption edge and not contributing to the
photoelectric absorption.
In ASCA observations Brandt et {\it al.} (1996)  observed an iron edge at 
$\sim 7.1 $ keV.
To explain the presence of this edge  they
discussed that a poorly ionized torus could be present around the 
neutron star, finding the relative covering fraction of $\sim 94 \%$
when the absorbed flux was $\sim 1.5 \times 10^{-9}$ erg cm$^{-2}$ s$^{-1}$
(low state). When the absorbed flux was 
$\sim 2.4 \times 10^{-8}$ erg cm$^{-2}$ s$^{-1}$ (high state),
the partial covering and the edge at 7 keV were no longer
detected. Their measures  of the galactic  N$_H$, $\sim 1.77  \times 
10^{22}$ cm$^{-2}$, and the  partial covering N$_H$, 
$\sim 1.5 \times 10^{24}$ cm$^{-2}$, are in agreement with our results. 
Shirey et {\it al.} (1999a)  observed a feature at 10 keV, present in the 
flaring branch during the high state of the source.   
In our observation the absorbed flux is   
$\sim 2 \times 10^{-8}$ erg cm$^{-2}$ s$^{-1}$ (high state), 
 and the energy of the edge is high, similar
to the value obtained by Shirey et {\it al.} (1999a).     
  
The torus of matter around Cir X--1  
could be a deformation of the outer region of the accretion disk due 
to the  tidal interaction between the two stars of the binary system and to
the high mass transfer from the  donor star. 
When Cir X--1 is in a low state the torus is poorly ionized, and it 
 partially absorbs the spectrum, and an edge from neutral 
iron and a fluorescence line at 6.4 keV are present. 
When Cir X--1 is in a high state the torus could be photoionized,
it is not necessary
to use a partial covering to fit the data,    
an emission  line at 6.7 keV is present  and the absorption
edge is shifted to higher energies.
The angle between the jet and the line
of sight is probably $>70^{\circ}$ (Fender et {\it al.}, 1998).
Assuming that the direction of the jet is 
perpendicular to the plane of the disk then the angle between the plane of the
disk and the line of sight is  $<30^{\circ}$. This could support the 
hypothesis that the line of sight intercepts  the torus.

In the hypothesis that the thickness of the torus is much
less than $d$, its distance from the source, we can estimate a  constraint  
on the distance of the torus from the neutron star  $d<L/N\xi$
(Reynolds \&
Fabian, 1995), where L is the unabsorbed luminosity, N is the column 
density of photoionized matter, and $\xi$ measures the ionization level.
In order to have Fe XXIII the ionization parameter is
$\xi \sim 100$ erg cm s$^{-1}$ (Turner et {\it al.}, 1992). For $L \sim 2 
\times 10^{38}$ erg s$^{-1}$ and N, the column density of photoionized matter,
$\sim 1.5 \times 10^{24}$ cm$^{-2}$ we obtain $d < 10^{12}$ cm. 
Considering that the minimum distance between the two stars of the binary has
an upper limit of  $\sim 6 \times 10^{11}$ cm 
(Murdin et   {\it al.}, 1980) we have only a weak constraint 
that the torus is within the system.

\section{Conclusions}

We analyzed data from a BeppoSAX observation of Cir X--1 performed in
1998 August 1. During the observation a flare lasting 6000 s was present.
 The energy spectrum is described
by a Comptonized  model, with a seed-photon temperature of $\sim 0.4$ keV and
electron temperature of  $\sim 0.9$. A hard 
power-law component is present in the spectra out of the flare.  
The radius of the seed-photon Wien distribution  is $\sim 150$  km.
The comptonized spectrum is produced in a relatively cool ($\sim 1$ keV) 
region of moderate optical depth ($\tau \sim 17$ for a spherical geometry).
The power-law  component, probably produced by scattering on high 
energy electrons in a jet,  has a photon 
index of $\sim 3$.  
We observe an absorption edge at $\sim 8.4$ keV,  probably originating
from Fe XXIII. Its optical depth requires  N$_H \sim 1.5 \times 10^{24}$ 
cm$^{-2}$, probably due to a highly ionized torus around the accretion disk.
In the spectra out of flare we  observe an iron emission line at 6.7 keV, 
with  equivalent widths $\sim 40-100$ eV whose energy indicates a high 
ionization level in agreement with that of the iron edge. 

\acknowledgments
This work was supported by the Italian Space Agency (ASI), by the Ministero
della Ricerca Scientifica e Tecnologica (MURST).

\clearpage

\section*{TABLE}
\begin{table}[th]
\begin{center}
\footnotesize
\caption{ \footnotesize In the first column we report 
the label number associated to  the intervals in which we divided 
our observation,
in the second column the start and end times (in second) for each interval
referred to  August 1 15:59:54.5 UT, the start time of MECS.
The average hardness ratio is the count rate in the energy 
band 3--7 keV divided by the count rate in 1--3 keV.  
The fluxes are in units of $10^{-8}$ ergs cm$^{-2}$ s$^{-1}$. The luminosity  
is in units of   $10^{38}$ ergs  $s^{-1}$, assuming a distance of the source 
of 5.5 kpc (Case \& Bhattacharya, 1998).  The 7th column indicates 
the intervals where the LECS data are available.}
\label{tab1}
\vskip 0.5cm
\begin{tabular}{c|c|c|c|c|c|c}

\tableline  
\tableline
Interval &Start--End time (s) & Average hard  & Flux &Flux &Luminosity& 
LECS data\\  
 & &  color & 2.5--25 keV& 0.1--100 keV &0.1--100 keV & \\
\tableline 
1 &    0--3300    & $0.61 \pm 0.04$  &         1.51&2.49&0.94&No\\
2 & 3700--4819    & $0.62_{-0.01}^{+0.04}$ &   2.00 &3.25&1.23&No\\
3 & 4820--5814    & $0.73 \pm 0.03$ &          2.88 &4.35&1.65&No\\
4 & 5815--6300    & $0.80 \pm 0.02$ &          3.27 &4.80&1.82&No \\
5 & 7500--9500    & $0.61_{-0.01}^{+0.04}$&     2.00 &3.26&1.24&No \\
6 & 9511--34322    & $0.53 \pm 0.04$ &          1.24 &2.24&0.85&Yes\\
7 &34326--58831    & $0.57 \pm 0.04$ &          1.35  &2.34&0.89&Yes\\
8 &58832--69829    & $0.60 \pm 0.02$ &          1.62  &2.69&1.02&No\\
\tableline
\end{tabular}
\end{center}
\end{table} 

\clearpage

\begin{table}[th]
\begin{center}
\footnotesize
\caption{ \footnotesize Results of the fit of Cir X--1 spectra during the 
flare in the energy 
band 0.1--100 keV, with a comptonized spectrum modeled by Comptt, a power law
and an absorption edge. Uncertainties are at 90\% confidence level for a 
single parameter. $kT_0$ is the temperature of the seed photons for the 
Comptonization, $kT_e$ is the electron temperature, $\tau$ is the optical 
depth 
of the scattering cloud using a spherical geometry, R$_W$ is the Wien radius
of the seed photons  in km and  f$_{bol}$ is the intrinsic flux of
the comptonized model in unit of ergs cm$^{-2}$ s$^{-1}$. 
The power law 
normalization is in units of photons keV$^{-1}$ cm$^{-2}$ s$^{-1}$ at 1 keV. 
F--Test
indicates the probability of chance improvement when the power law is 
included in the spectral model.}

\vskip 0.5cm
\begin{tabular}{l|c|c|c|c|c}

\tableline  
\tableline

Interval                                    & 1 &2 &3 &4 &5\\                                          
\tableline                                 &    &  &  & &   \\
$N_{\rm H}$ $\rm (\times 10^{22}\;cm^{-2})$&1.740 (frozen)&  1.740 (frozen) & 1.740 (frozen) & 1.740 (frozen) & 1.740 (frozen)\\
 E$_{edge}$ (keV)                            &  $8.387 \pm 0.069$& $8.25 \pm 
0.14$ &  $8.38 \pm 0.29$ & $8.57 \pm 0.18$ & $7.83^{+0.19}_{-0.15}$\\

$\tau_{\rm edge}$                          & $0.609 \pm 0.079$&$0.73^{+0.21}_{-0.18}$ & $0.31 \pm 0.12$ & $0.50 \pm 0.16$ & $0.56^{+0.19}_{-0.15}$\\
$k T_0$ (keV)                              &  $0.397 \pm 0.024$ & $0.422 \pm
0.019$ &  $0.431 \pm 0.021$ &  $0.448^{+0.034}_{-0.028}$ & $0.400 \pm 0.030
$\\
$k T_{\rm e}$ (keV)                        &  $0.921^{+0.056}_{-0.050}$&
 $1.026^{+0.096}_{-0.063}$ & $1.043^{+0.051}_{-0.044}$ & $1.089^{+0.077}_{-0.072}$ & $0.901^{+0.078}_{-0.046}$\\
$\tau$                                     &   $16.6^{+1.9}_{-1.6}$ & $14.2^{+1.6}_{-1.8}$ & $15.9 \pm 1.3$ & $15.6^{+2.0}_{-1.7}$ & $17.4^{+2.0}_{-2.4}$\\
 N$_{\rm comp}$                            &  $34.1^{+3.2}_{-4.1} $ & 
$38.3^{+4.0}_{-4.6}$ & $43.3^{+3.2}_{-3.3}$ & $42.1^{+4.8}_{-4.3}$ & 
$45.3^{+4.8}_{-5.6}$\\

f$_{bol}$($\times 10^{-8}$)     & 5.5&6.2&8.9&9.4&7.2\\ 
R$_W$ (km)             & $140 \pm 20$ & $142 \pm 17$ & $151 \pm 17$ & 
                         $144 \pm 25$ &  $156 \pm 28$ \\

 Photon index                & $3.27^{+0.76}_{-0.83}$ & 3.30 (frozen) & 3.30 (frozen) &  3.30 (frozen) &  3.30 (frozen)  \\

Power--Law N                  &  $1.7^{+12.1}_{-1.6}$ & $< 1.14$ &$< 0.78$ &
$< 1.3$ & $1.61^{+0.85}_{-0.93}$ \\

$\chi^2$/d.o.f.                             & 231/234 & 238/259 & 224/200 & 228/252& 190/232\\ 
F--Test             & $\sim 0$ & 0.3 & 1 & 1 & $2 \times 10^{-3}$ \\
\tableline
\end{tabular}
\end{center}
\end{table}

\begin{table}[th]
\begin{center}
\footnotesize
\caption{ \footnotesize Results of the fit of Cir X--1 spectra out the 
flare in the energy 
band 0.1--100 keV, with a comptonized spectrum modeled by Comptt, a power law,
a gaussian emission line
and an absorption edge. Uncertainties are at 90$\%$ confidence level for a 
single parameter. Definitions are as in Table 2. 
$EQW_{\rm Fe}$ indicates the equivalent width of the gaussian emission line,
E$_{Fe}$ the centroid of the  emission iron line and I$_{Fe}$ is the 
normalization of the emission iron line in units of photons cm$^{-2}$ 
s$^{-1}$.}

\vskip 0.5cm
\begin{tabular}{l|c|c|c}

\tableline
\tableline

Interval                                   &6 &7 &8 \\
\tableline                                 & & & \\              
$N_{\rm H}$ $\rm (\times 10^{22}\;cm^{-2})$ & $1.727^{+0.113}_{-0.048}$ &
$1.737^{+0.181}_{-0.075}$  & 1.740 (frozen)\\
  E$_{edge}$ (keV)                               & $8.495 \pm 0.053$  & 
$8.477^{+0.076}_{-0.066}$ &$8.594^{+0.082}_{-0.073}$ \\
$\tau_{\rm edge}$                          & $1.02 \pm 0.13$  & $0.844^{+0.053}_{-0.167}$ & $0.73 \pm 0.11$\\
$k T_0$ (keV)                             & $0.378^{+0.010}_{-0.013}$ &
 $0.382^{+0.016}_{-0.024}$ &  $0.420^{+0.022}_{-0.015}$\\
$k T_{\rm e}$ (keV)                        & $ 0.839^{+0.025}_{-0.029}$ &
 $0.859^{+0.025}_{-0.056}$ & $0.911^{+0.039}_{-0.046}$\\
$\tau$                                     & $16.67^{+0.92}_{-0.76}$ & 
  $17.2^{+2.0}_{-1.3}$ & $16.0 ^{+1.4}_{-1.1}$\\
 N$_{\rm comp}$                            & $40.9^{+3.0}_{-2.3}$ & 
 $38.2^{+5.2}_{-3.3}$ & $36.3^{+2.4}_{-3.4}$\\
f$_{bol}$($\times 10^{-8}$)     &5.5&5.4&5.8\\
R$_W$ (km) &$161\pm 13$ &$152\pm 22$&$134\pm 16$ \\

  Photon index                    & $3.16^{+0.48}_{-0.39}$ & 
 $3.34^{+0.50}_{-0.43}$ & $3.37^{+0.50}_{-0.47}$\\
 Power--Law N                         & $1.5^{+4.2}_{-1.0}$ & 
$2.5^{+7.2}_{-1.8}$ & $2.8^{+8.5}_{-2.1}$\\
E$_{Fe}$ (kev)                            & $6.70 \pm 0.10$ & 
$6.82^{+0.14}_{-0.28}$ & $6.31^{+0.30}_{-0.61}$ \\
$\sigma_{\rm Fe}$ (keV)                    & $ < 0.24$  & 
 $ < 0.47$ & $0.50^{+0.51}_{-0.30}$\\
I$_{\rm Fe}$                               & $(1.34^{+0.60}_{-0.56}) \times 10^{-3}$ & $(2.8^{+7.0}_{-1.6}) \times 10^{-3}$ & $(5.7^{+13.6}_{-3.4}) \times 10^{-3}$\\

$EQW_{\rm Fe}$ (eV)                          & 46.1 &  100. & 82.2 \\

$\chi^2$/d.o.f.                             & 685/591 & 651/609 & 141/194\\ 
F--Test             & $\sim 0$  &$\sim 0$  &$\sim 0$  \\

\tableline
\end{tabular}
\end{center}
\end{table} 

\clearpage
 
\section*{FIGURE CAPTIONS}
\bigskip

\noindent
{\bf Figure 1}:  
Upper panel: Cir X--1 light curve in the energy band 
1.8--10 keV (MECS data);  the selected time intervals
are indicated  at the top of the panel.  
Lower  panel: The ratio of the count rate
in the energy  band 3--7 keV with respect to that in 1--3 keV. 
The bin time is 137 s.\\
{\bf Figure 2}: The color-color diagram of Cir X--1. 
The hard color is the ratio of the count rate  in the energy band 7--10 keV
with respect to that in the band 3--7 keV, while the soft color is 
the ratio (3--7 keV)/(1--3 keV). The bin time is 137 s. 
The main variation visible in the diagram is due to the observed flare.\\ 
{\bf Figure 3}: 
Energy spectra (0.1--100 keV) of Cir X--1 for intervals 6 (out of the flare,
left side) and 4 (peak of the flare, right side).
Data and the corresponding best fit model (see Tab. 2 and 3) are shown 
in the upper panels, residuals in units of $\sigma$ with respect to the 
best fit model are shown in the lower panels.\\
{\bf Figure 4}: 
Left Panel: Unfolded spectrum and the best fit model, shown in this figure 
as a solid line, for interval 6. The single components of the model are 
also shown, namely the  {\it Comptt} (dashed line), Power Law (dotted line)
and iron emission line (dot-dashed line).  The absorption edge at 
8.4 keV is visible.
Right Panel: Unfolded spectrum and the best fit model, shown as a solid line,
for interval 4.  The absorption edge at 8.4 keV is visible. \\
{\bf Figure 5}: 
Seed-photon temperature (upper panel) and electron temperature (lower panel) 
of the comptonized spectrum as a function of the time, where the time values 
are the centers of each interval (see Tab. 1). \\
{\bf Figure 6}: 
The electron temperature of the comptonizing region as a function of the 
seed-photon temperature. The solid line is a linear fit, with a slope of
$\sim 3.2$.

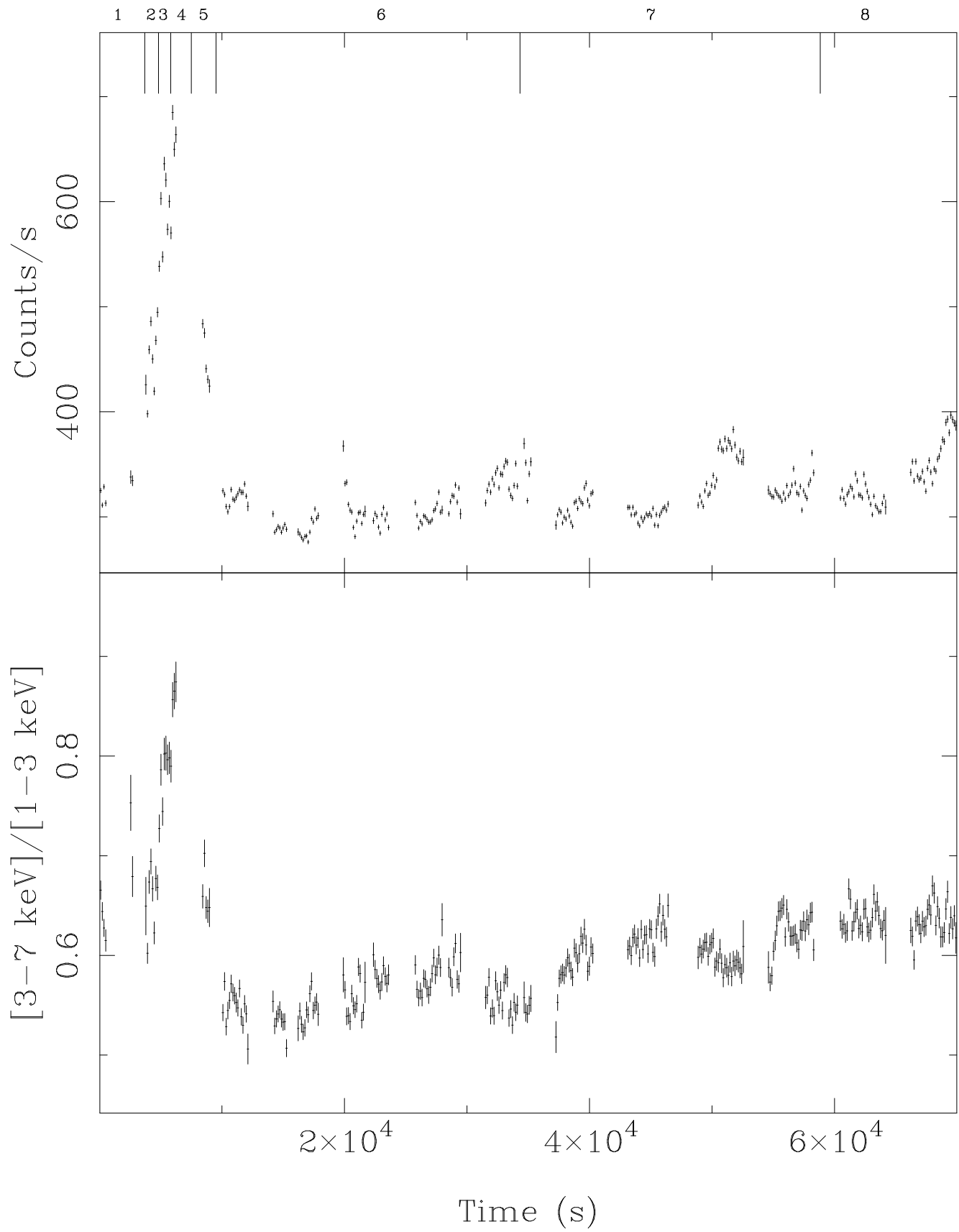
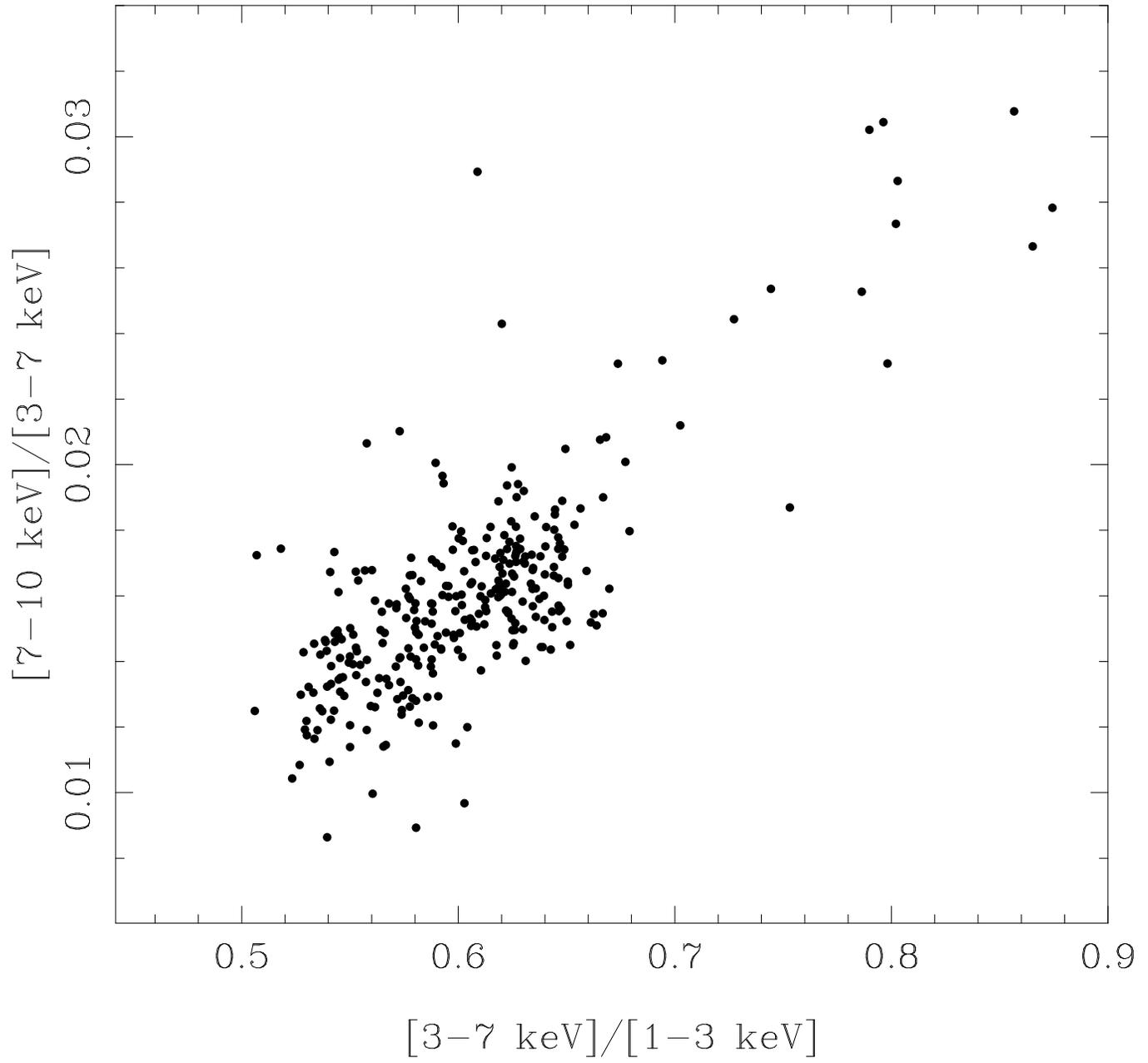
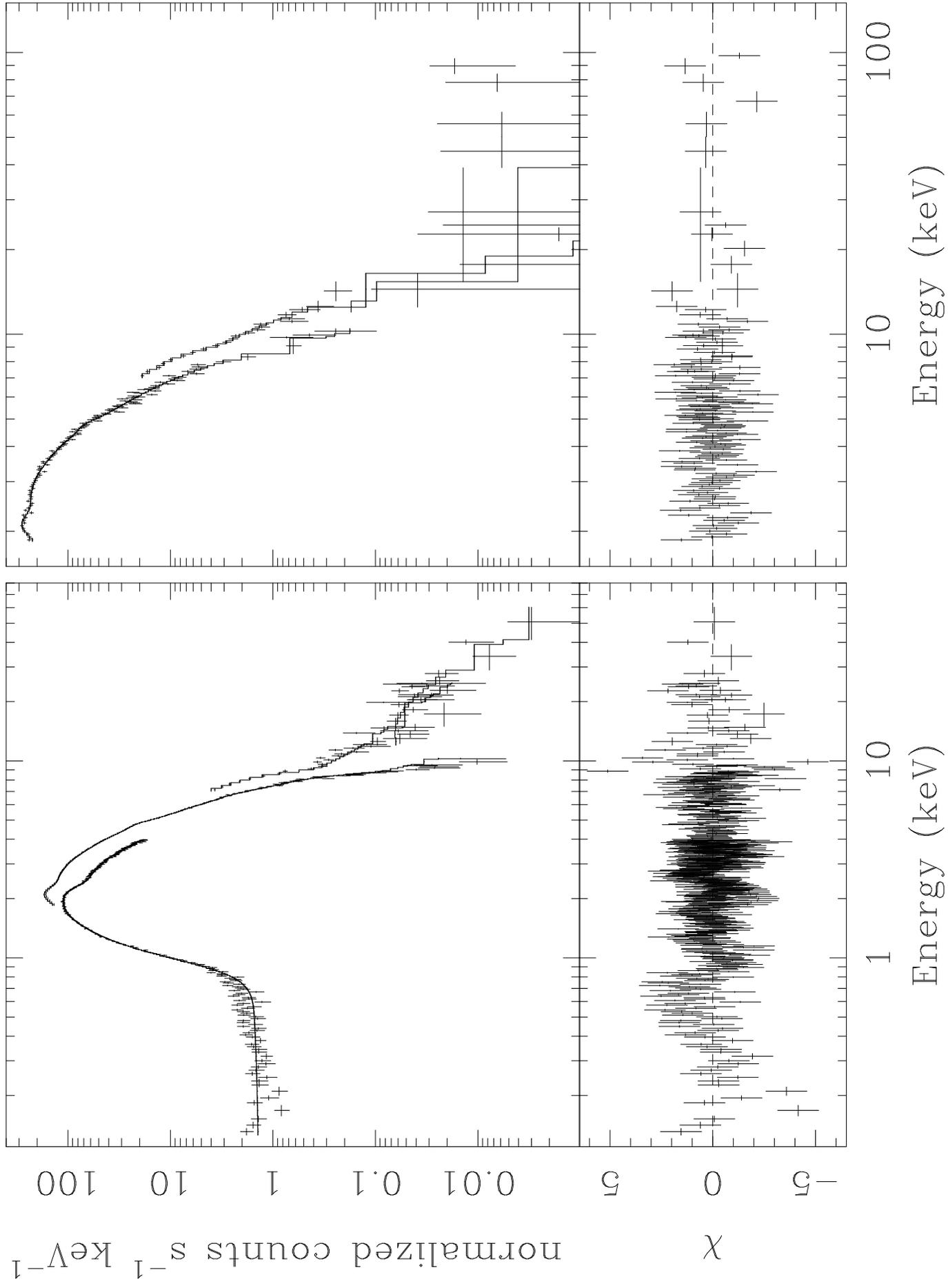
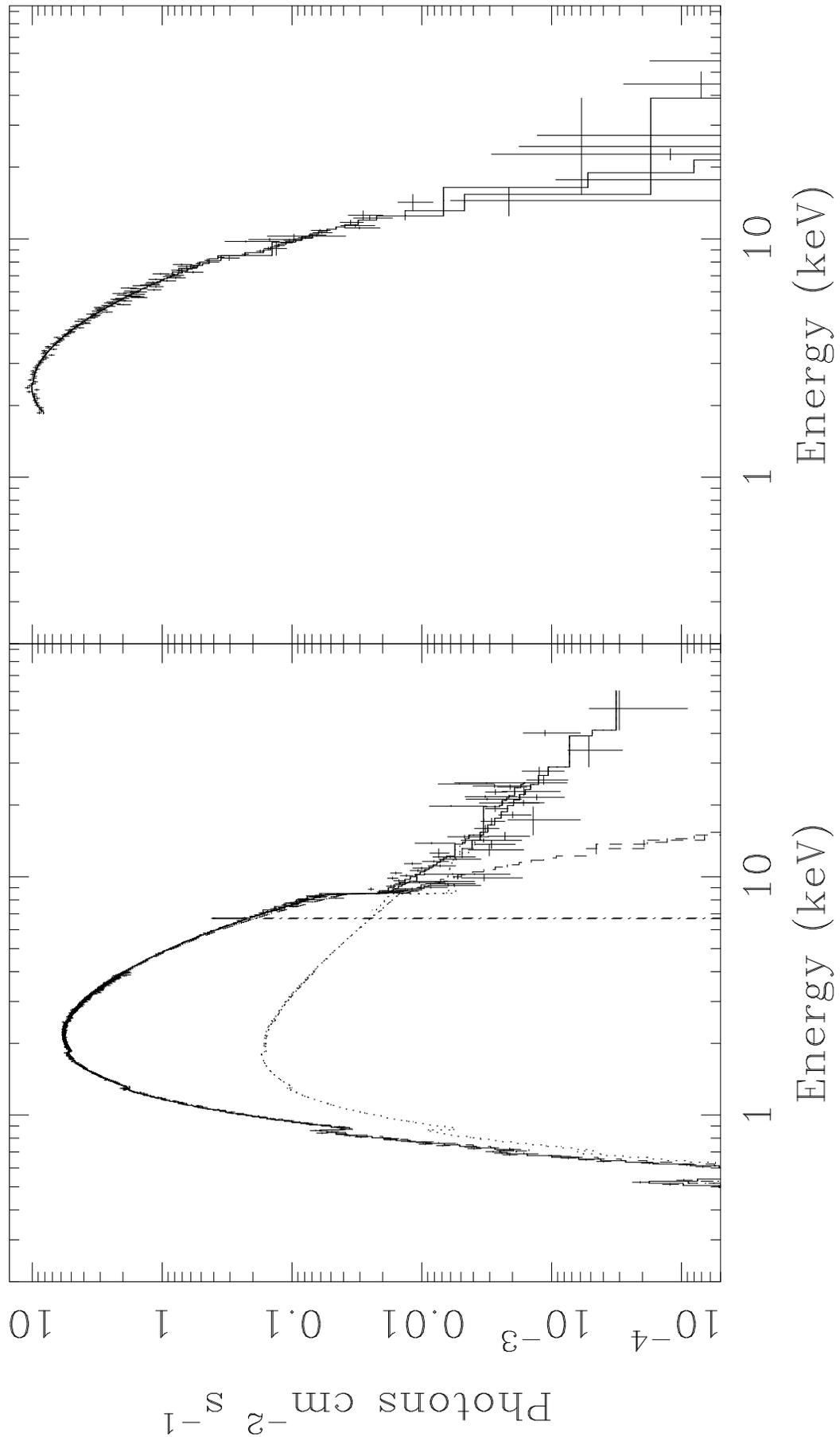
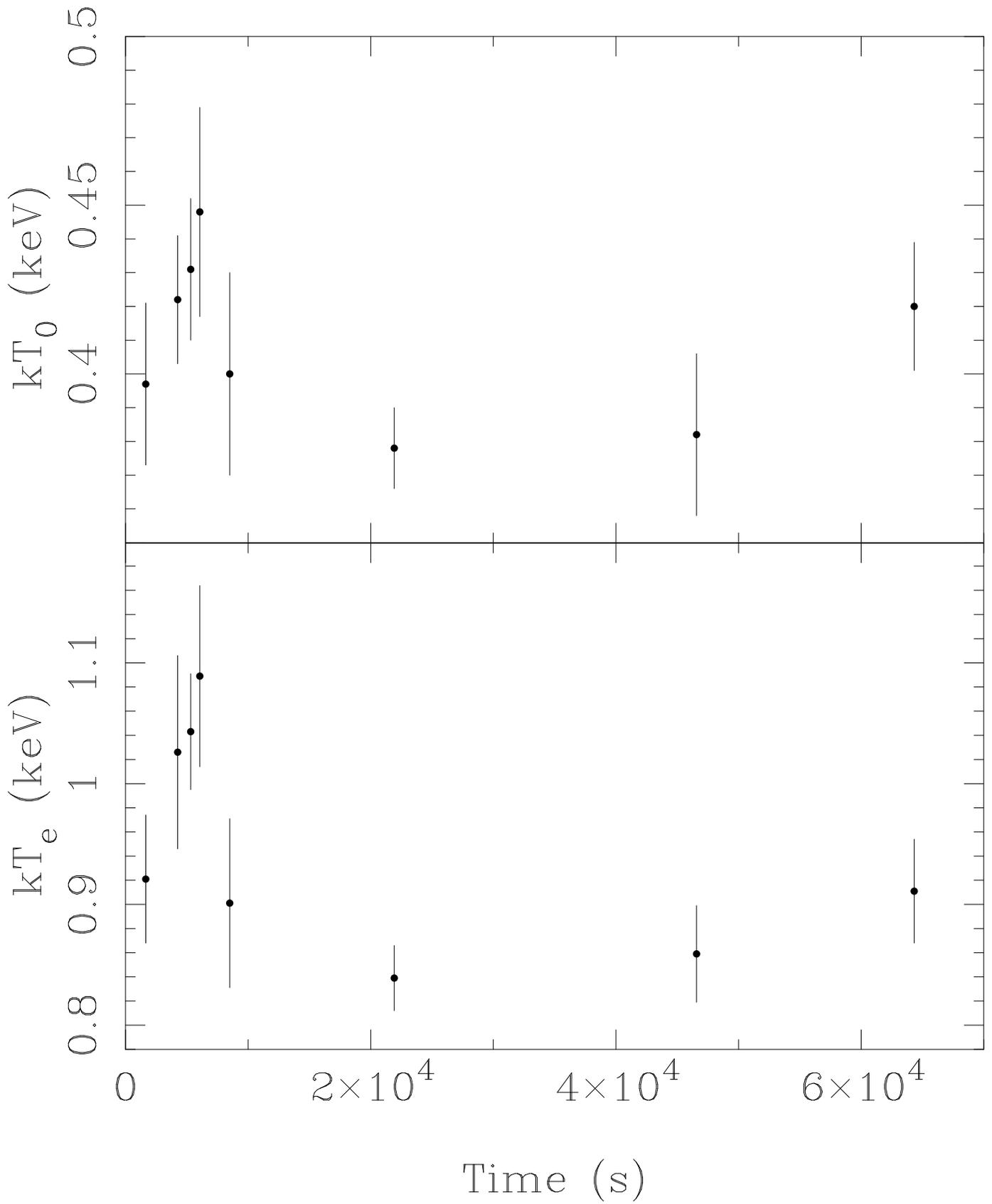
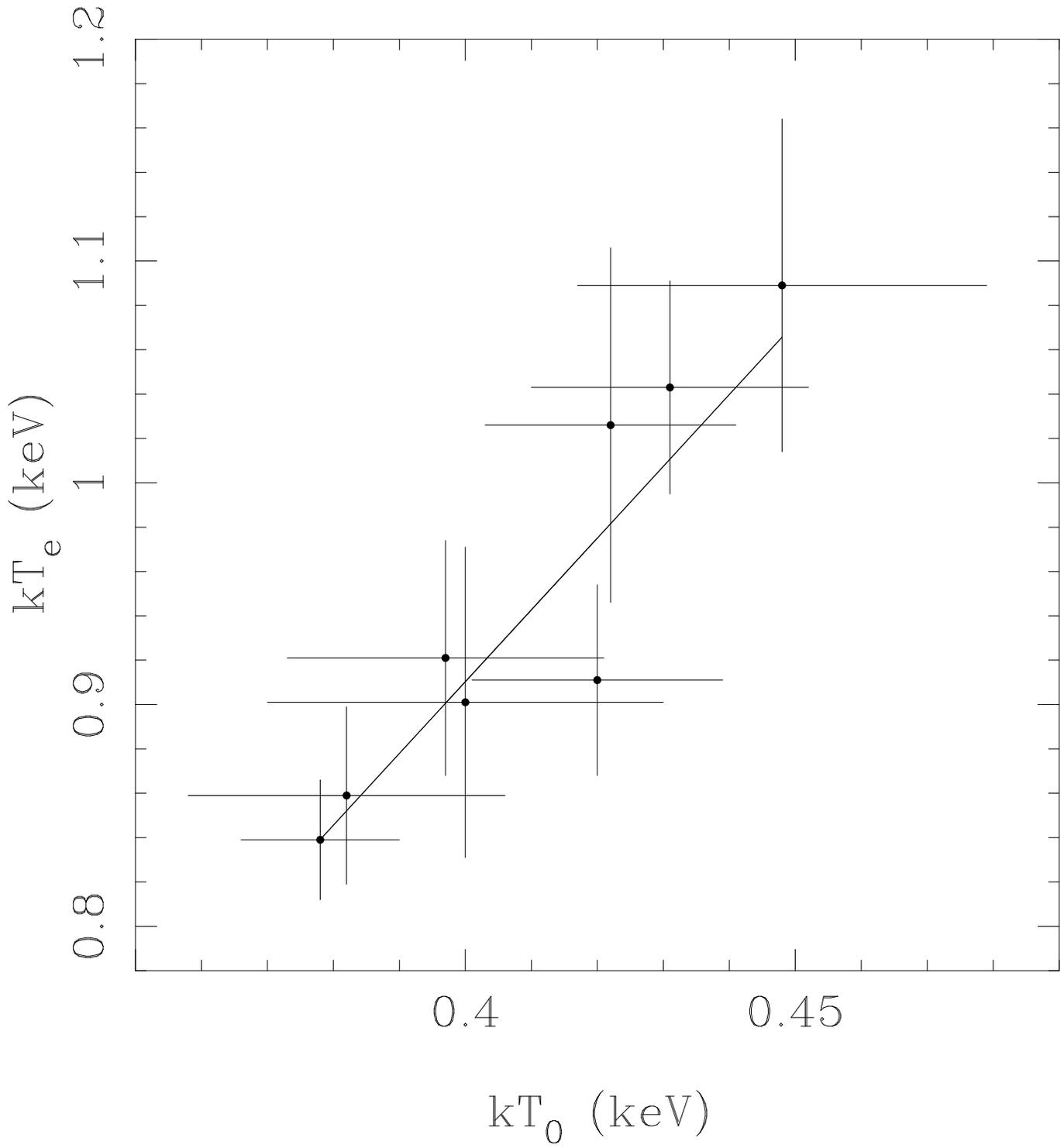
\end{document}